\documentclass{iau}

\usepackage{amsmath}
\usepackage{graphicx}
\usepackage{multirow}
\usepackage{orcidlink}

\newcommand{\nbody}{\textit{N}-body }

\begin{document}

\lefttitle{Cho et al.}
\righttitle{Modeling TDEs \& Compact Object Plunges in NSCs}

\jnlPage{1}{7}
\jnlDoiYr{2025}
\doival{10.1017/xxxxx}

\aopheadtitle{Proceedings IAU Symposium}
\editors{C. Sterken,  J. Hearnshaw \&  D. Valls-Gabaud, eds.}

\title{Modeling Tidal Disruption Events and Compact Object Plunges in Nuclear Star Clusters}

\author{Philip Cho\orcidlink{0009-0003-7328-2504}$^1$, 
        Kai Wu\orcidlink{0000-0003-0349-0079}$^1$, 
        Francesco Flammini Dotti\orcidlink{0000-0002-8881-3078}$^{2,3}$, 
        Taras Panamarev\orcidlink{0000-0002-1090-4463}$^4$, 
        Rainer Spurzem\orcidlink{0000-0003-2264-7203}$^{1,5,6}$}

\affiliation{$^1$Astronomisches Rechen-Institut, Heidelberg University, 69120 Heidelberg, Germany}
\affiliation{$^2$Department of Physics, New York University Abu Dhabi, PO Box 129188 Abu Dhabi, UAE}
\affiliation{$^3$Center for Astrophysics and Space Science (CASS), New York University Abu Dhabi, PO Box 129188 Abu Dhabi, UAE}
\affiliation{$^4$Fesenkov Astrophysical Institute, Observatory 23, 050020 Almaty, Kazakhstan}
\affiliation{$^5$National Astronomical Observatories, Chinese Academy of Sciences, 100101 Beijing, China}
\affiliation{$^6$Kavli Institute for Astronomy and Astrophysics, Peking University, 100871 Beijing, China
}

\begin{abstract}
We study tidal disruption events (TDEs) and compact object inspirals in nuclear star clusters (NSCs) hosting a central supermassive black hole (SMBH), focusing on their role in SMBH growth. Using the STARDISK version of the direct \nbody code \texttt{NBODY6++GPU}, we perform pilot simulations with two improved models: one for mass fallback from TDEs and another for compact object plunges based on orbital decay timescales. Our results show that mass accretion via TDEs peaks within the first 2 Myr and decreases more rapidly for higher initial SMBH masses, with roughly half the disrupted stellar debris being accreted. Compact object accretion is confined mostly to orbits with pericenters between 4 and 27 Schwarzschild radii and is suppressed by an order of magnitude when inspiral criteria are applied.
\end{abstract}


\maketitle

\section{Introduction}
Nuclear star clusters (NSCs) frequently reside in the centers of galaxies, often coexisting with a central supermassive black hole (SMBH) \citep{2020A&ARv..28....4N, 2009MNRAS.397.2148G, 2010RvMP...82.3121G}. These dense stellar systems provide a rich environment for complex dynamical interactions that may result in two significant classes of astrophysical phenomena: tidal disruption events (TDEs) \citep{1988Natur.333..523R, 2021ARA&A..59...21G} and extreme mass ratio inspirals (EMRIs). TDEs occur when stars pass sufficiently close to the SMBH to be torn apart by tidal forces, producing luminous flares as the bound stellar debris accretes onto the SMBH, a plausible channel for SMBH growth, particularly in gas-deficient environments \citep{1989ApJ...346L..13E}. EMRIs arise when compact objects such as stellar-mass compact objects gradually spiral into the SMBH via the emission of gravitational waves \citep{EMRI_2015}. These unique sources of gravitational waves provide precise information about the spacetime geometry near the SMBH and allow for measurements of its spin \citep{2018LRR....21....4A}.

This study presents an improved modeling framework for both TDEs and EMRIs using high-resolution direct N-body simulations. We employ the NBODY6++GPU code, which is designed for large particle number simulations using supercomputers \citep{s41115-023-00018-w}. We introduce two new models: (1) a refined TDE model that accurately computes the fraction of bound debris following stellar disruptions, and (2) an EMRI model that estimates inspiral timescales of compact objects and their eventual merger with the central SMBH. We also present pilot simulations using these models and analyze their implications for SMBH growth in NSCs.

\section{Methodology}
We use the \texttt{NBODY6++GPU} code with its specialized \textit{STARDISK} branch, optimized for nuclear star cluster simulations\footnote{\url{https://github.com/nbody6ppgpu/Nbody6PPGPU-beijing/tree/stardisk-dev}}. In the original \textit{STARDISK} implementation \citep{2019MNRAS.484.3279P}, the SMBH is modeled as a fixed external potential, and a tidal radius is a fixed value, defined as the product of the SMBH's  influence radius and an accretion parameter. Stars passing this radius are removed from the simulation and assumed to be fully accreted. While this approach provides a simplified framework for modeling TDEs, it neglects the fact that only a fraction of the stellar debris becomes bound and accretes onto the SMBH. Compact objects should also be treated separately, as they go through relativistic inspirals.

To improve this treatment, we developed a model that computes the bound mass fraction based on the star's orbital parameters at the time of disruption. Specifically, we use a classification scheme for TDEs based on the orbital eccentricity and the penetration factor 
$\beta$ (the ratio of the tidal radius to the pericenter distance), dividing events into five types according to the relationship between the star’s eccentricity and two critical eccentricities \citep{2018ApJ...855..129H}. The fallback rate of debris is computed using a top-hat distribution with a uniform density assumption, leading to an analytical estimate of the bound mass.

To model EMRIs, we select compact objects within half the influence radius of the SMBH, ensuring that their dynamics are primarily dominated by the SMBH. For each, we compute the merger time via the Peters formula \citep{1963PhRv..131..435P, 1964PhRv..136.1224P} and if this time becomes shorter than the Keplerian orbital timescale, the object is assumed to merge with the SMBH.

\subsection{Initial Conditions}
For the pilot simulations, we initialize a post-starburst NSC consisting of 128,000 single stars, without primordial binaries for simplicity. For the initial particle distribution of the system, the Plummer model is used \citep{1911MNRAS..71..460P, 1987MNRAS.224...13D}. The mass of star particles ranges from 0.08$M_{\odot}$ to 150$M_{\odot}$ and are distributed by the Kroupa initial mass function (IMF) \citep{Kroupa_2001}. Three initial SMBH masses (5\%, 10\%, and 20\% of the total cluster mass) are used to investigate their impact on TDE demographics. The system is evolved for $\sim$110 Myr.

In the EMRI simulation, a special cluster of only stellar-mass black holes is used to enhance EMRI statistics. These black holes span a mass range between  8 and 100~$\mathrm{M_{\odot}}$, and the central SMBH is initialized with 10\% of the total cluster mass. This simulation lasted for $\sim 4 \mathrm{Myr}$.

\section{Results}
\begin{figure}[h]
    \centering
    \includegraphics[width=1.0\columnwidth]{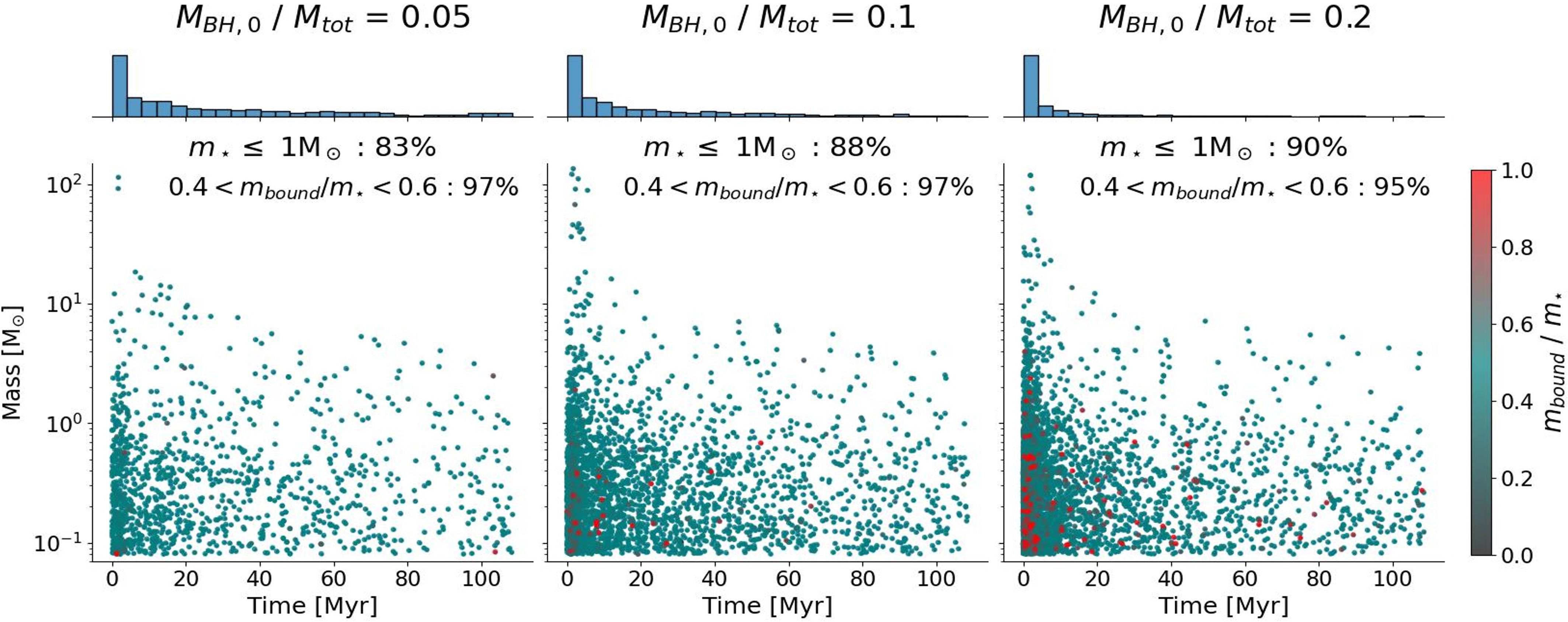}
    \caption{TDE distribution by time, color-coded by the bound debris mass fraction. Normalized histograms of distributions are shown on top of the plots.}
    \label{fig:1}
\end{figure}

Figure \autoref{fig:1} shows the temporal distribution of TDEs, which indicates that the majority occur early in the simulation. In all three cases with varying initial SMBH masses, the number of TDEs peaks within the first 20 Myr. This concentration shifts toward even earlier times as the initial SMBH mass increases. Most of the disrupted stars are low-mass main-sequence stars, which contribute more than 80\% of all TDEs. More than 95\% of these TDEs yield a bound debris mass close to 50\%. However, for higher initial SMBH masses, a broader range of bound mass fractions is observed, suggesting deeper penetrations and more eccentric orbits. Most TDEs are classified as marginally eccentric or marginally hyperbolic, with the bound mass fraction distributions clustering around the parabolic regime. A larger initial SMBH spreads the TDE type distribution, yielding more events with higher bound mass fractions.

\begin{figure}[h]
    \centering
    \includegraphics[width=1.0\columnwidth]{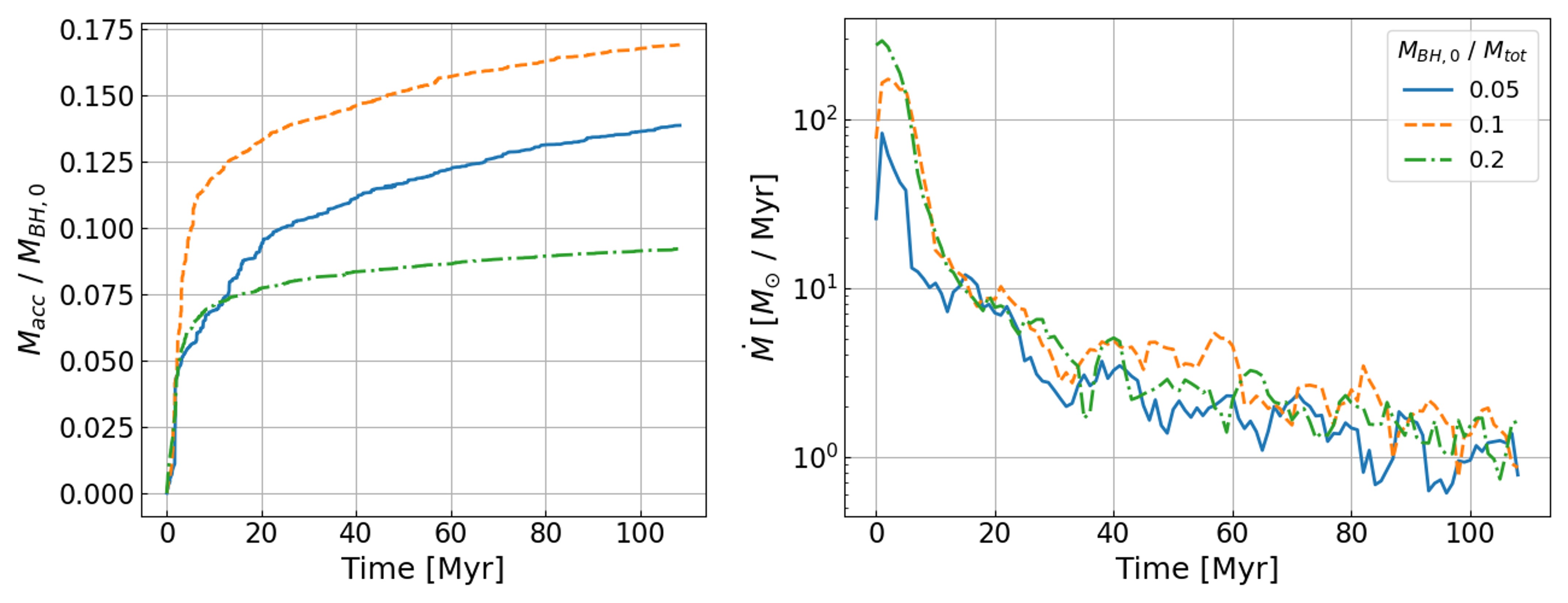}
    \caption{Evolution of SMBH mass growth (left panel) and mass accretion rates (right panel).} 
    \label{fig:2}
\end{figure}

Figure \autoref{fig:2} shows that the SMBH accretion rate mirrors the TDE distribution in time, peaking within the first 2 Myr and then decaying by nearly two orders of magnitude. This suggests that a significant fraction of SMBH mass growth may come from TDEs shortly after a starburst. The simulation with an initial SMBH mass equal to 10\% of the total cluster mass shows a consistently higher accretion rate beyond the initial peak. Cumulatively, the mass growth of the most massive SMBH remains below 10\%, whereas the SMBHs with 5\% and 10\% initial masses grow by approximately 15\%. This indicates a nonlinear relationship between initial SMBH mass and TDE-driven growth, likely due to rapid depletion of loss-cone stars in more massive systems.

The EMRI simulation using a cluster of stellar-mass black holes provides insight into compact object accretion processes. We analyze the pericenter distances, $r_p$, of all accreted objects and compare them with thresholds derived from previous studies. Most compact objects fall within the range $4r_s < r_p < 27r_s$, where $r_s$ is the Schwarzschild radius. However, several outliers are identified beyond this range. These objects typically have long merger times and should not have been accreted so early. Their presence indicates limitations in the current inspiral criterion and motivates further refinement.

\begin{figure}[h]
    \centering
    \includegraphics[width=0.6\columnwidth]{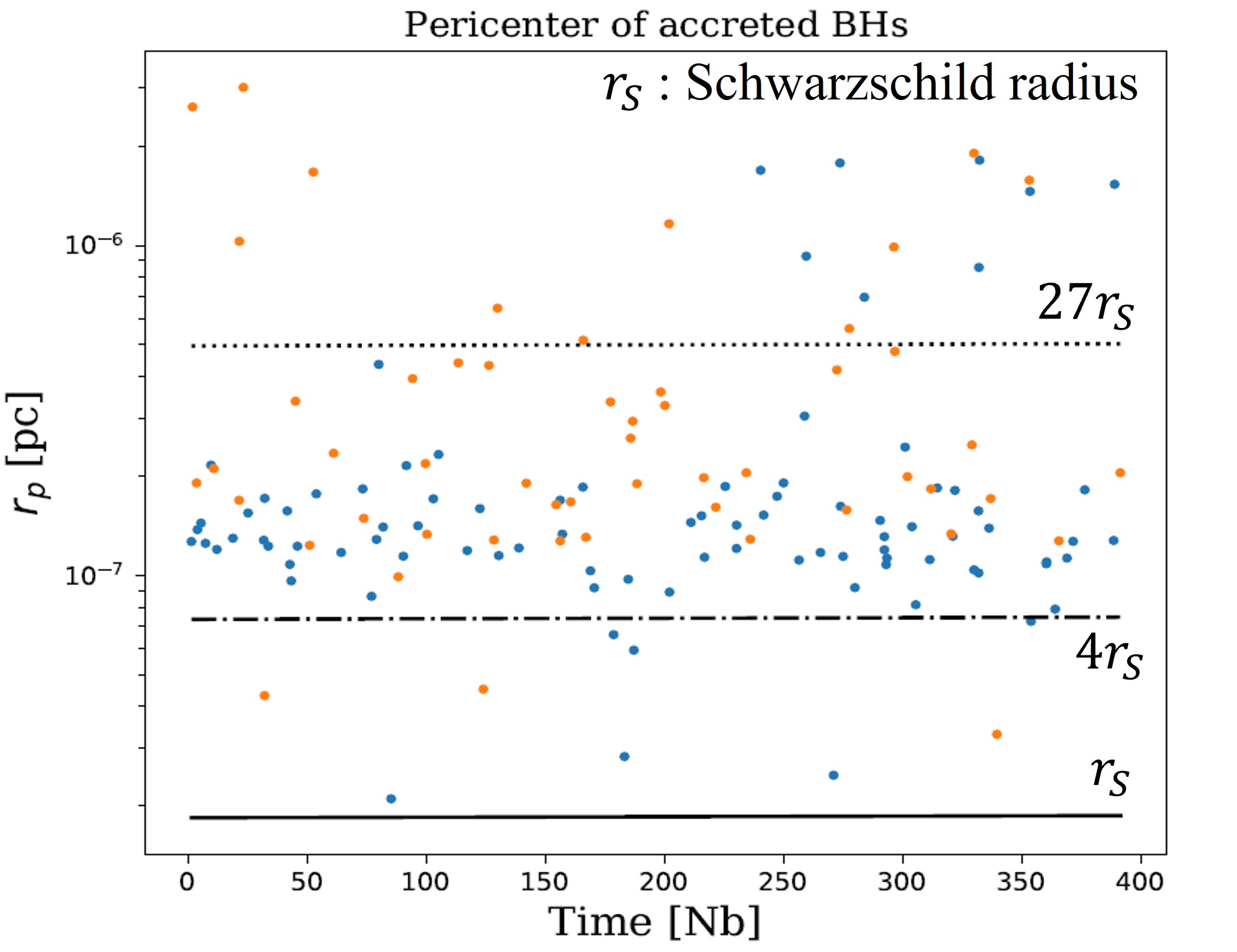}
    \caption{Pericenter distances of stellar-mass black holes at the time of accretion. Blue dots represent those with eccentric orbits, and orange dots represent those with hyperbolic orbits. The simulation time of 400NB is approximately 4Myr} %\kw{maybe mention the nb time to Myr conversion factor here}} 
    \label{fig:3}
\end{figure}

\section{Conclusions and Outlook}
Our simulation results reveal that the vast majority of TDEs are initiated by low-mass main-sequence stars in the early stages following a starburst. These events typically result in approximately half of the disrupted stellar mass being bound and accreted by the SMBH. For systems with more massive initial SMBHs, distribution of bound mass fractions is broader, reflecting more deeply penetrating orbits. Accretion rates across all models peak within the first few Myr and decline sharply thereafter, with the intermediate-mass SMBH model sustaining the highest average accretion over time. The EMRI model captures the expected distribution of pericenter distances for stellar-mass black holes, though it also highlights areas needing improvement, as some compact objects are accreted prematurely due to oversimplified inspiral criteria. Overall, our findings support the role of TDEs and EMRIs as influential mechanisms in SMBH growth, particularly in the post-starburst phase of NSC evolution.

These preliminary results lay the foundation for the \texttt{DRAGON-III} project on star cluster simulations, incorporating both models in a unified framework with a larger number of particles and longer evolutionary timescales.

\end{document}